# Synthesis and structural analysis of Mo/B periodical multilayer X-ray mirrors for beyond extreme ultraviolet optics

*Oleksiy V. Penkov[a*], Igor A. Kopylets[b], Valeriy V. Kondratenko[b], and Mahdi Khadem[c]*

*[a] ZJU-UIUC Institute, International Campus, Zhejiang University, Haining 314400, China*

*[b] National Technical University "KhPI," Kharkiv 61002 Ukraine*

*[c] Department of Mechanical Engineering, Yonsei University, Seoul 03722 Korea*

**Abstract:** The synthesis and structural analysis of Mo/B periodical multilayer X-ray mirrors (PMMs) for beyond extreme ultraviolet (BEUV) optics was performed. The PMMs were deposited by a combination of pulsed DC and radio frequency (RF) magnetron sputtering. The structure was analyzed by high-resolution transmission electron microscopy, X-ray photoelectron spectroscopy, and grazing incidence X-ray reflectometry. The formation of 0.35 nm-thick interlayers comprised of a mixture of molybdenum borides was observed at the Mo/B interfaces. Furthermore, a low interface roughness of 0.3-0.4 nm was reported. The temperature of the substrate increased due to the increase in the sputtering power. This resulted in an increase in the thickness of the interlayers and the interface roughness; subsequently, the optical properties of the PMM deteriorated. Theoretical calculations were performed based on the real structure of the PMM to predict the reflectivity at a working wavelength of 6.7 nm. The reflectivity of approximately 49% was two times higher than that of the conventional $B_4C$-based BEUV mirrors. Based on our study results, it can be concluded that the as-synthesized PMMs will perform better than the traditional mirrors and can be effectively used for the development of the next generation of BEUV lithography.

*Corresponding author email: oleksiypenkov@intl.zju.edu.cn

## 1. Introduction

There is a continuous trend of scaling down the manufacturing processes in the computer industry to increase the operating frequency and decrease the power consumption of microprocessors [1, 2]. Thus, increasing the resolution of the lithography equipment used in the industry and the corresponding decrease in the working wavelength is required [3]. Recently, the resolution has been increased by replacing the deep ultraviolet (DUV) wafer scanners with novel devices that use extreme ultraviolet (EUV) radiation ($\lambda = 13.5$ nm) [4]. A shortening of the working wavelength to 6.7 nm and the use of beyond extreme ultraviolet (BEUV) radiation will further enhance the performance of microprocessors [4, 5].

The weak reflection of EUV and BEUV radiation by solids necessitates advanced reflective optical devices like periodical multilayer X-ray mirrors (PMMs). PMMs are artificial Bragg crystals with alternate layers of "light" and "heavy" materials; their periodicity is approximately half of the working wavelength in the case of the close to normal incidence [6, 7]. The first step of the design of a high-efficient PMM includes the selection of a material pair. The choice is grounded in the optical properties of materials in the given range of wavelengths. For example, Mo/Si PMMs are mostly utilized for the wavelength of 13.5 nm (EUV), and Mo/B PMM are optimal for 6.7 nm (BEUV) [8-10]. The next step is choosing the thickness of the individual layers and their number. This step is usually based on computer simulations [11]. For instance, for BEUV radiation and the incidence angle 5° off normal, the thickness of B (light) and Mo (heavy) layers should be around 2.25 and 1.12 nm, respectively. Supplementary Figure S1 shows the example of the calculated reflectivity of such Mo/B PMM as a function of the number of pairs. It could be seen that the theoretical reflectivity rapidly growth and exceed 55% while the number of pairs increased up to 250.

Further increasing of the number of pairs to 400 would increase the theoretical reflectivity to ~58%. For the reliable design of PMMs, it is crucial to characterize the real nanostructure of PMMs. Various imperfections, such as interface roughness and intermixing between layers, will reduce the interfaces' sharpness and deteriorate the overall reflectivity [12, 13].

The material-based design of PMM for the wavelength range of 2-12 nm was performed by Claude et al. [10]. It was shown that Mo/Si PMMs are not suitable for the BEUV range due to a combination of optical properties. Instead, according to their calculations, Mo/B PMMs should exhibit high reflectivities at 6.6–11.5 nm wavelengths. But a plethora of studies has been conducted on Mo/$B_4C$, Pd/$B_4C$, and La/$B_4C$ PMMs that utilize boron carbide ($B_4C$) as a substitute for B [14-18]. In contrast, Mo/B PMMs have been rarely investigated despite their superior theoretical performance compared to their competitors. The disparity in the volume of research is attributed to complexity of the deposition of the high-quality Mo/B PMMs. Furthermore, the experimentally measured reflectivity of Mo/B PMMs was less than 10 % [10]. The marked difference is attributed to the significant interface roughness of approximately 0.65 nm and the low number of pairs.

The complexity in the manufacturing of Mo/B PMMs is primarily attributed to the dielectric nature of boron. Long-time magnetron sputtering of the dielectric target might cause instability of the deposition process due to the electrical charging of surfaces. In the case of $B_4C$, the behavior of the deposition process was more predictable. Thus, boron in the PMMs was mostly replaced by the less effective $B_4C$.

The manufacturing of high-performance PMMs involves a lot more than the alternating stacking of two materials. In particular, preserving periodicity is crucial for maintaining Bragg's diffraction condition. The periodicity implies uniformity of the layer's thickness and structure of interfaces along the whole multilayer stack. The quality of the periodicity is mostly defined by the deposition rate's time stability and substrate temperature [19].

Besides, the precise control of the layers' structure and composition, interface roughness, and spatial uniformity is required. The nanostructure of the PMMs has a critical impact on the optical properties that can be controlled by a thorough understanding of the relation between the structure and deposition conditions. For example, La/B PMMs exhibited a low reflectivity due to the intermixing of the layers and the high interface roughness [16]. However, knowledge about the physical mechanisms of the intermixing allowed further improvement of the deposition process. In particular, the nitridation of the interfaces between La and B was proposed [20]. The nitridation of La suppresses the intermixing of Mo and B, thereby increasing the optical contrast between La and B. The EUV reflectivity of the PMMs is also significantly increased by nitridation. However, periodicity errors occurred due to the technical complexity of the selective nitridation. These errors did not allow reaching the theoretical reflectivity.

The structure of the Mo/B PMMs remains practically unknown due to the complexities mentioned above in the synthesis. The research on the optical properties of Mo/B PMMs did not include a detailed structural analysis [10]. This study demonstrated the synthesis and structural analysis of high-quality Mo/B PMMs. The effect of the deposition parameters on the structure and optical properties of the PMMs was also investigated. The structural analysis was performed to predict the theoretical reflectivity of the real Mo/B BEUV mirrors at a working wavelength of 6.7 nm.

## 2. Experimental details

Coatings were deposited by magnetron sputtering in a custom-built deposition system (Infovion, Korea) with a base pressure of ~$10^{-4}$ Pa. The deposition setup consisted of a vertical cylindrical vacuum chamber with magnetrons mounted on the top (Supplementary Figure S2). Mo and B targets were installed on the magnetrons. A substrate holder was mounted below the magnetrons and was moved from one magnetron to another by a robotic arm to deposit alternate layers of Mo and B. The target-to-substrate distance was 100 mm. A shutter was used to control the deposition time.

A pulsed DC (40 kHz, 50 mA) power source was used to sputter Mo; the power density Mo was 0.2 W/cm$^2$. Radiofrequency (RF) power source was used to sputter B. The power of the B magnetron was ranged from 280 to 330 W; the power density was in the range of 3.5-4.1 W/cm$^2$.

The deposition rates for each power were preliminarily calibrated. The thickness of the Mo and B layers was controlled by adjusting the exposition time. Sputtering was performed under an Ar atmosphere at 0.26 Pa. The working pressure was measured by a Baratron® gauge and was controlled by an automatic throttle valve. A thermo-resistive sensor was directly attached to the Cu substrate holder by thermal grease to monitor the substrate temperature during deposition.

Mo/B PMMs with different Mo and B thickness were deposited on polished Si wafers. The B nominal thickness was ranged from 2 to 18 nm. The nominal thickness of the Mo layers was 1.5–9 nm. The number of periods in PMMs was varied from 16 to 300 to maintain the same total thickness of coatings. The bottom and top of each PMM comprised a ~7 nm-thick Mo layer and ~14 nm-thick B layer.

Grazing incidence X-ray reflectometry (GIXR, DRON-3M, Bourevestnik, Russia) was performed with Cu-Kα radiation (λ= 0.154 nm, 60 kV/30 mA) in the θ–2θ geometry (Bragg-Brentano configuration). A Si (220) single-crystal monochromator was used to separate the Cu-K$_{\alpha 1}$ line and collimate a primary beam with a divergence of 0.015°. The slit size was 100 μm. A combination of GIXR and computer simulation was utilized to determine the thickness, density, and the interface roughness of the individual layers of the PMMs. The computer simulation implemented the recursive computation method of XRR, which entirely considers the effects of dynamic scattering and absorption [12, 13]. The embedded database of the scattering factors based on the paper of Henke et al. [21] was used. The calculation was performed using X-Ray Calc software; the calculation method and theoretical background are described elsewhere in detail [11]. The calculated GIXR curves were fitted to the measured GIXR curves by adjusting the parameters of the computer model that indicated the chemical composition, thickness, density, and interface roughness of each layer.

The structure of the PMMs was investigated by cross-sectional high-resolution transmission electron microscopy (HRTEM, JEOL JEM-ARM200F, Japan). The cross-sections of the samples for HRTEM were prepared by a focused ion beam (FIB, JEOL JIB-4601F, Japan).

The chemical composition of the PMMs was evaluated by X-ray photoelectron spectroscopy (XPS, Thermo Fisher Scientific Mono). The XPS peaks were deconvoluted with the Gauss-Lorentz functions using the Levenberg–Marquardt algorithm. The inelastic electron scattering background was approximated using the active Shirley method [22].

The curvature of the substrate (base length = 20 mm) was measured by a 3D stylus profilometer (Dektak, Bruker, USA) after the deposition of the PMMs. It was subsequently utilized to calculate the biaxial film stress using the modified Stoney's equation [23]. The Young's modulus and Poisson's ratio of the Si substrates were 130 GPa and 0.28, respectively.

### 3. Results and discussion

A series of Mo/B PMMs was synthesized for the detailed investigation of the growth mechanism. The example of Mo/B PMM's structure is shown in a cross-sectional HRTEM image (Figure 1a,b), and it is described in Table 1. The coating started from a 7 nm thick Mo sublayer deposited onto Si substrate. Then, 16 pairs of Mo and B with a thickness of ~ 11.2 and 2.88 nm were deposited on the top of the sublayer. 12 nm thick B layer covered the PMM. The dark and bright stripes in the HRTEM image (Figure 1(a)) corresponded to Mo and B layers, respectively. The amorphous structure of the Mo and B layers was revealed by Fourier transformation (inset of Figure 1a). The formation of the amorphous Mo layer was attributed to the solid-state amorphization mechanism [24, 25, 26].

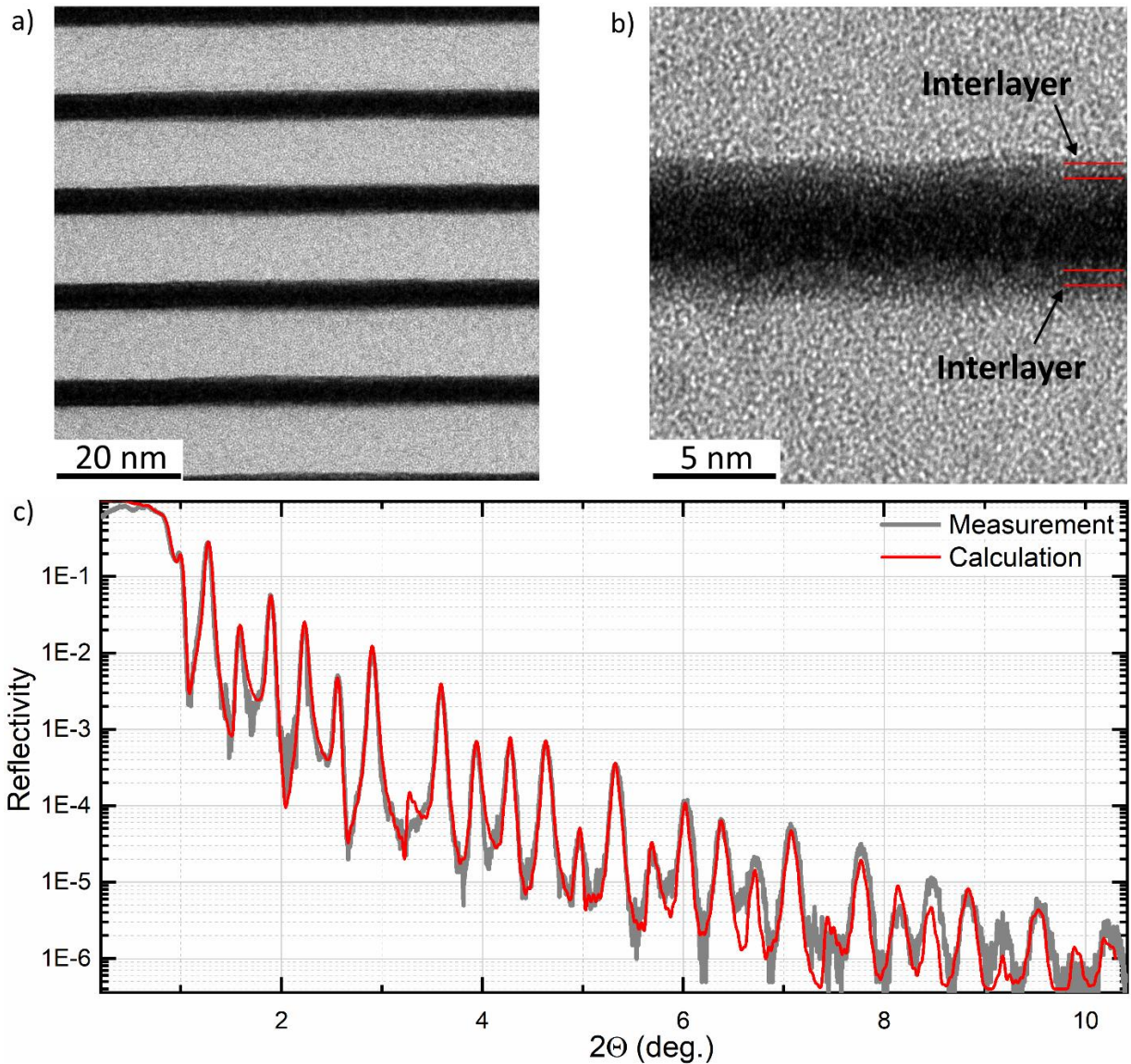

**Figure 1.** Structure of the PMM deposited with B power of 300W having thicknesses of the Mo and B layers of 5 and 11 nm, respectively. a) Cross-sectional HRTEM image. b) Magnified image of the central region. c) GIXR fitting. The model structure is shown in Table 1.

**Table 1.** Parameters of the realistic model with interlayers for the GIXR simulation in Fig. 1c (power of the B magnetron is 300 W).

| Stacks | N | Layers | | | |
|---|---|---|---|---|---|
| | | Material | Thickness, nm | Roughness, nm | Density, g/cm$^3$ |
| Top | 1 | Boron | 12 | 0.8 | 1.8 |
| Main | 16 | $MoB_2$ | 0.43 | 0.3 | 7.0 |
| | | Molybdenum | 2.88 | 0.3 | 10 |
| | | $MoB_2$ | 0.48 | 0.28 | 7.0 |
| | | Boron | 11.3 | 0.23 | 2.37 |
| Sublayer | 1 | Molybdenum | 7 | 0.5 | 10.2 |
| Substrate | - | Silicon | $\infty$ | 0.5 | 3.2 |

The gray strips (Figure 1(b)) indicated the symmetrical interlayers between the Mo and B layers. The HRTEM analysis revealed that the thickness of the interlayers was approximately 0.5 nm. Since the thickness was low, the HRTEM analysis did not indicate whether the intermediate contrast was caused by the intermixing between Mo and B or by the roughness of the interface. To clarify this point, the GIXR simulation was performed with different model structures. The best fit between the theoretical and experimental GIXR curves (Figure 1(c)) was obtained when 0.35 nm-thick interlayers were introduced into the model (Table 1). In this model, the interlayers were consisted of a mixture of Mo and B atoms, and the atomic ratio of components was similar to that of $MoB_2$. The best fitted density of the interlayers in the model was 7.0 g/cm$^{3,}$ which was ~3% lower than the density of bulk molybdenum boride. Such a decrease in density is typical for amorphous materials.

The thickness of the interlayers obtained from the GIXR simulation was slightly lower than that obtained from the HRTEM analysis. This was attributed to the small tilt of the cross-sectional specimen during the measurement. The thickness of the layers was significantly lower than the total thickness of the sample in the projection direction. Therefore, the tilt of the cross-sectional specimen induced the extension of shadows on the image. Moreover, it was difficult to precisely measure the thickness of the interlayers from the HRTEM images due to the interface roughness. Therefore, the data obtained from the GIXR simulation were more trustworthy than those obtained from the HRTEM analysis.

For further investigation of the deposition mechanism and formation of interlayers, another series of PMMs was prepared. During the deposition of this series, deposition rates were fixed, and deposition times were varied to obtain PMMs with different layer's thickness. The power of B magnetron was 300 W. In the first group of PMMs, the Mo thickness was 1.5 nm, and the nominal thickness of B was varied from ~2 to ~18 nm. In the second group, Mo thickness was varied from ~1.5 to ~9 nm. Since all these coatings were deposited under the same conditions, and the deposition time was the only variable, a similar structure of the interface was expected [19, 27, 28]. GIXR data reveleted that these PMMs had the structure of the Mo/B interfaces similar to the one shown in Table 1. The real thicknesses of the Mo and B layers were obtained from the GIXR data and plotted as functions of the deposition time (Figure 2).

There was a linear increase in the thickness of the individual layers with an increase in deposition time. The linear approximation of the growth curves in Figure 2 demonstrated that the y-intercept of the Mo growth curve laid above zero, and that of the B growth curve laid below zero. Therefore, both the growth curves were shifted from their natural positions. This behavior indicated the formation of interlayers during the deposition. The growth curves in other layered structures such as Sc/Si, Mo/Si, and W/B$_4$C exhibited a similar behavior [19, 27, 29]. The negative shift of the B growth curve indicated the diffusion or "consumption" of B into the Mo layer, causing the formation of interlayers at the interface. The "swelling" of the Mo layer and the positive shift of the Mo growth curve was attributed to the formation of the Mo-B compounds.

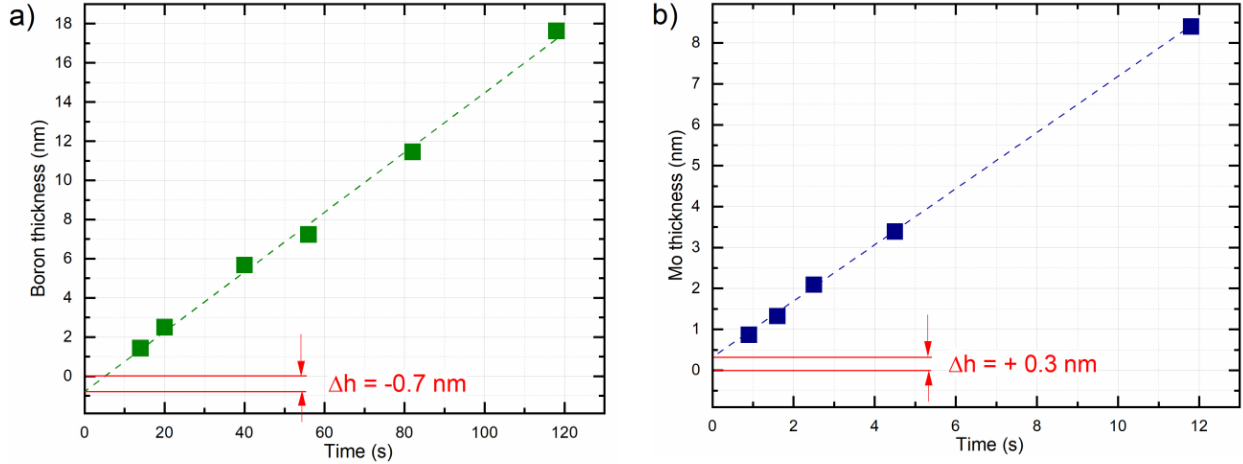

**Figure 2.** Growth curves for the power of the B magnetron of 300 W. Variation in the thickness of the (a) boron and (b) Mo layers with the deposition time.

The chemical composition of the PMMs was evaluated by XPS depth profiling to confirm the intermixing between B and Mo. The estimated etching rate was about 0.02 nm/sec. The variation in Mo and B concentrations with the etching time is shown in Figure 3(a). The left part of the graph corresponded to the thick B layer at the surface. The periodical structure was reached after 500 s of etching; however, the oscillations due to the periodicity of the structure were not defined due to the low thickness of the layers in the PMM. Figures 3(b)–(d) show the Mo 3d and B 1s peaks corresponding to the etching times of 126, 990 s, and 1090 s (Vertical lines 0, 1, and 2 in Figure 3(a)). These etching times represented the local minimum and maximum of the profile.

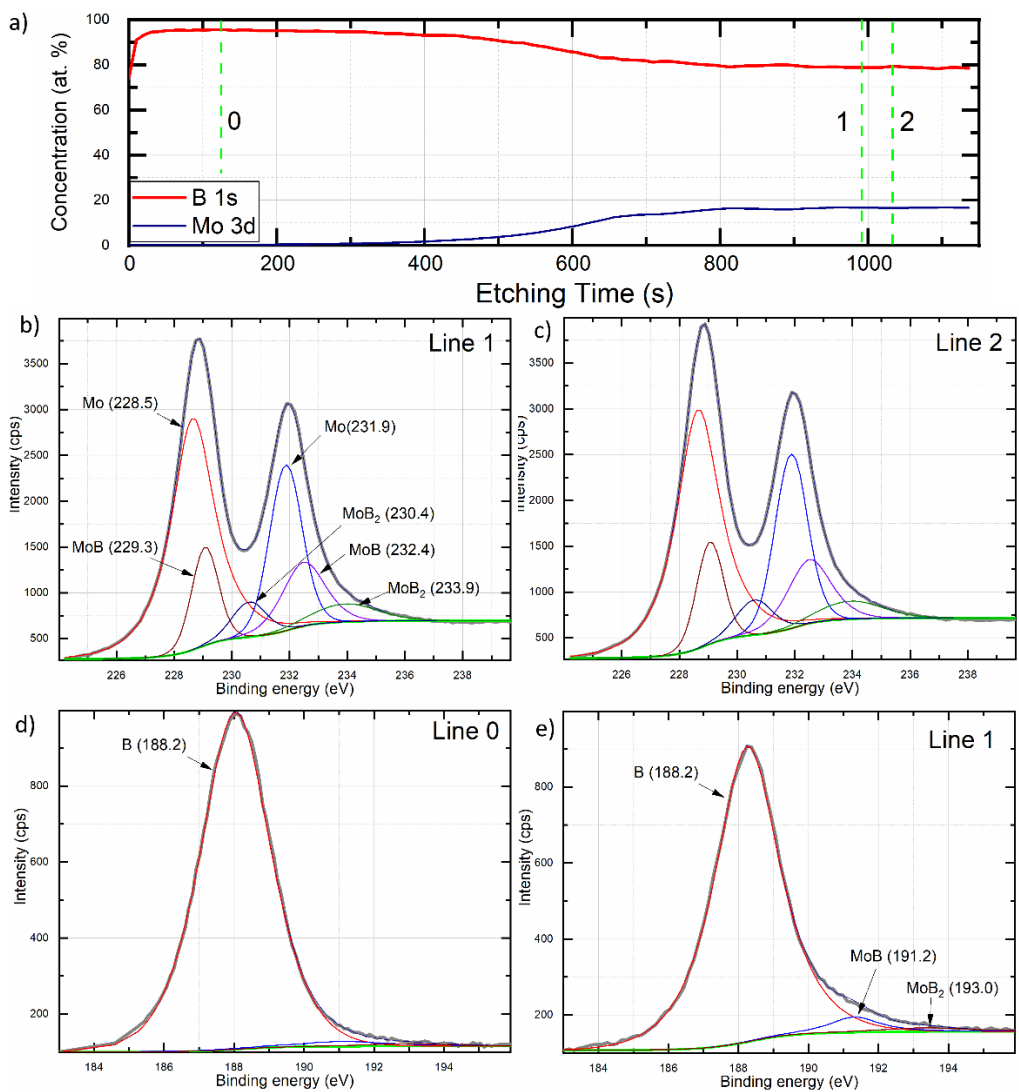

**Figure 3.** XPS Data for the Mo/B PMM, where the thickness of the Mo and B layers was 2.2 nm and 1.1 nm, respectively. The power of the B magnetron was 300 W. a) Variation in the concentration of Mo and B with the etching time. b-c) Deconvolution of the Mo 3d peak corresponding to the etching times of 990 s and 1090 s. d-e) Deconvolution of the B 1s peak corresponding to the etching times of 126 s and 990 s.

The deconvolution of the Mo and B peaks indicated the presence of pure B and Mo in conjunction with the Mo-B compounds. The spectra (Figure 3(b)–(c)) showed that the primary Mo $3d_{5/2}$ and Mo $3d_{3/2}$ peaks were located at 228.5 eV and 231.9 eV, respectively. Several secondary peaks with intensities lower than those of the peaks of pure Mo were observed at 229.3 eV, 230.4 eV, 232.4 eV, and 233.9 eV. The presence of these secondary peaks indicated the chemical bonding between Mo and B to form molybdenum boride (MoB) and molybdenum diboride (MoB₂) [30-32]. The deconvolution of the B 1s peak indicated the presence of a central peak at 188.3 eV that corresponded to pure B (Figure 3(d)). The presence of several secondary peaks was attributed to the chemical bonding between Mo and B to form MoB and MoB₂.

The paired comparison of Mo and B at different etching times showed a slight variation in the amount of the Mo-B compounds (Figures 3(b)–(c), 3(d), and 3(e)). The amount of pure Mo at Line 1 was lower than that at Line 2. The XPS data could not be quantified further due to the limited depth resolution and the preferential sputtering [33]. However, the presence of the XPS peaks corresponding to pure boron, molybdenum, and the molybdenum borides confirmed the existence of interlayers between the pure Mo and B layers.

Then, the effect of B magnetron source power on the structure of Mo/B interfaces was further investigated. The measured deposition rate of B was 0.03 nm/s and 0.067 nm/s at 200 W and 300 W, respectively. The measured deposition rate of Mo was approximately 0.2 nm/sec at 50 mA. The deposition of 300 Mo/B pairs was a time-consuming process; therefore, the B magnetron power was increased to overcome the low sputtering yield of B. However, this resulted in a significant rise in the substrate temperature (Fig. 4(a)). When the power increased from 280 W to 330 W, the final temperature of the substrate increased from 78 °C to 120 °C. This negatively affected the structure of the PMM by growing the interface roughness and the thickness of the interlayers (Figure 4(b), Supplementary Tables S1-S3). The variation of the temperature during the deposition increased the non-uniformity of the PMMs due to the continuous increase in the thickness of the interlayers.

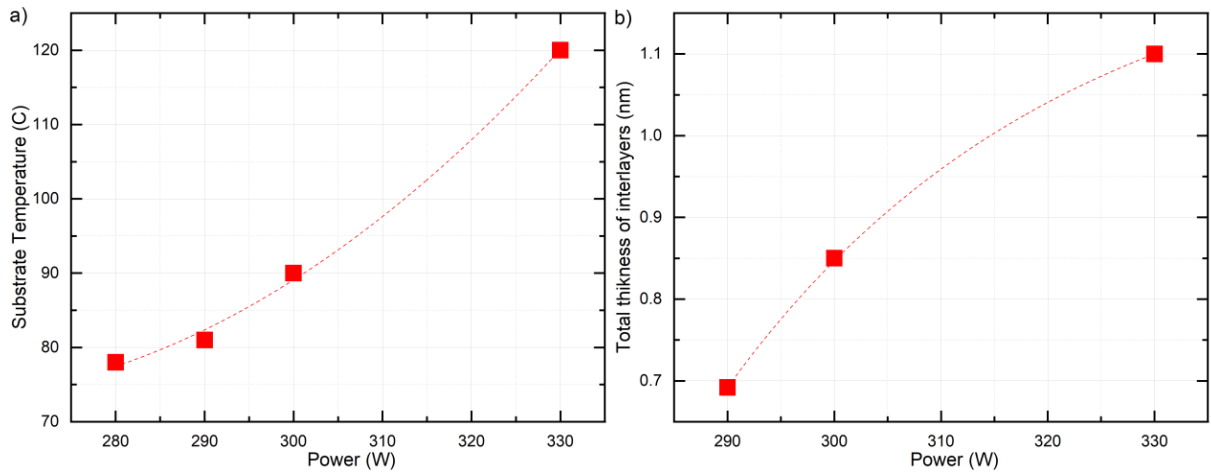

**Figure 4.** Effect of power on B magnetron on a) the final substrate temperature; b) the total average thickness of the interlayers in PMMs (based on GIXR simulation).

Knowledge about the growth mechanism and structure of interlayers between Mo and B allowed propper design the x-ray mirror for the wavelength of ~6.7 nm. The PMM was deposited with the power of B magnetron set at 280W. The nominal thickness of Mo and B was 1.2 and 2.2 nm, respectively. The number of Mo/B pairs was 300. The GIXR for this PMM is shown in Figure 5, and the corresponding GIXR model is described in Table 2.

The GIXR curve in Figure 5 is more suitable for comparing the best fits for the 4- and 2-layer models than the curve in Figure 1c. The most significant difference was observed for the 3rd diffraction order. The intensity of the 3rd order diffraction peak in the 2-layer model was three times higher than that in the measured curves. The fitting of the calculated and measured curves was prevented by any physically significant variation in the parameters of the 2-layer model.

**Table 2.** Parameters of the realistic model with interlayers for the GIXR simulation in Figure 5. The power of the B magnetron was 280 W

| Stacks | N | Layers | | | |
| --- | --- | --- | --- | --- | --- |
| | | Material | Thickness, nm | Roughness, nm | Density, g/cm$^3$ |
| Top | 1 | Boron | 8 | 0.8 | 2.37 |
| Main | 300 | MoB$_2$ | 0.35 | 0.3 | 7.0 |
| | | Molybdenum | 0.76 | 0.3 | 10.2 |
| | | MoB$_2$ | 0.35 | 0.3 | 7.0 |
| | | Boron | 1.89 | 0.23 | 2.37 |
| Sublayer | 1 | Molybdenum | 5 | 0.5 | 10.2 |
| Substrate | - | Silicon | ∞ | 0.5 | 3.2 |

Then, the expected performance of the PMMs at the wavelength of 6.7 nm was evaluated, and the effect of the structural imperfections was compared. It was shown in [10] that a multilayer X-ray mirror with the ideal periodical structure comprises Mo and B layers with sharp interfaces and exhibits the highest reflectivity at a working wavelength of 6.7 nm. The simulation results indicated that the peak reflectivity of such an ideal PMM was 60 % (Figure 6). The reflectivity of the real BEUV mirrors was heavily dependent on the interface roughness. The data obtained from the HRTEM and GIXR analysis revealed that the roughness of the relatively smooth interfaces of the real Mo/B PMMs was approximately 0.3 nm. The introduction of this interface roughness into the ideal model lowered the reflectivity to 52 %. The presence of the interlayers blurred the interfaces; thus, the reflectivity of the real Mo/B PMMs decreased further in comparison to that of the ideal model.

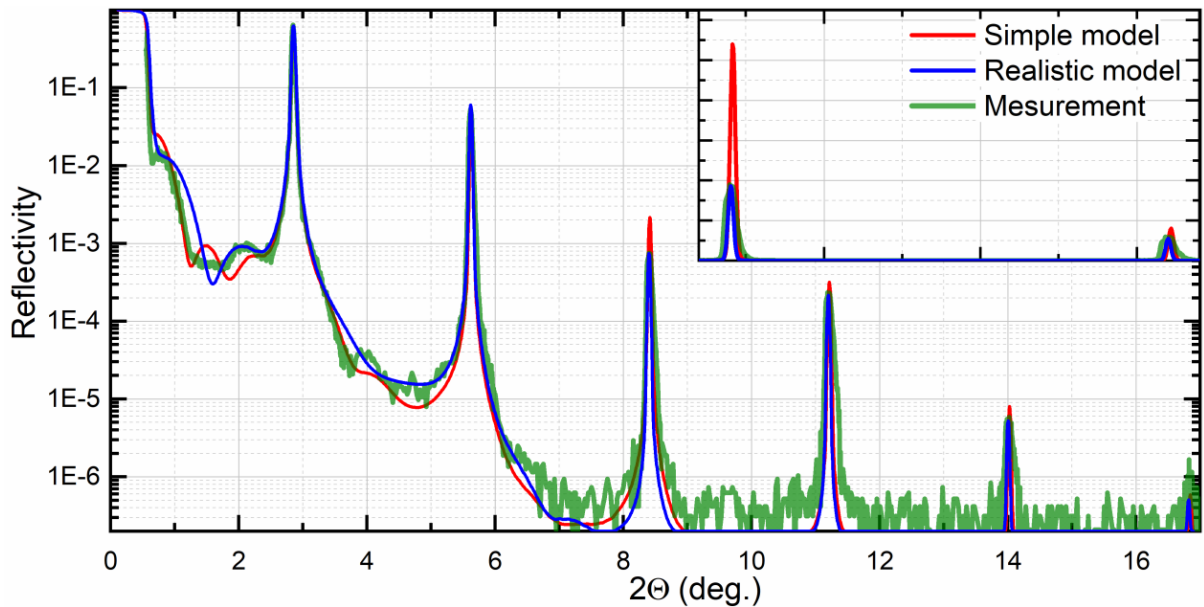

**Figure 5**. GIXR fitting. An example of two best-fitted model curves for the Mo/B PMM. The total number of Mo/B pairs was 300 and the power of the B magnetron was 280 W. The simple model consisted only of Mo and B layers. The realistic model included 0.4 nm-thick $MoB_2$ interlayers as shown in Table 2. The inset shows the 3rd and 4th diffraction orders on a linear

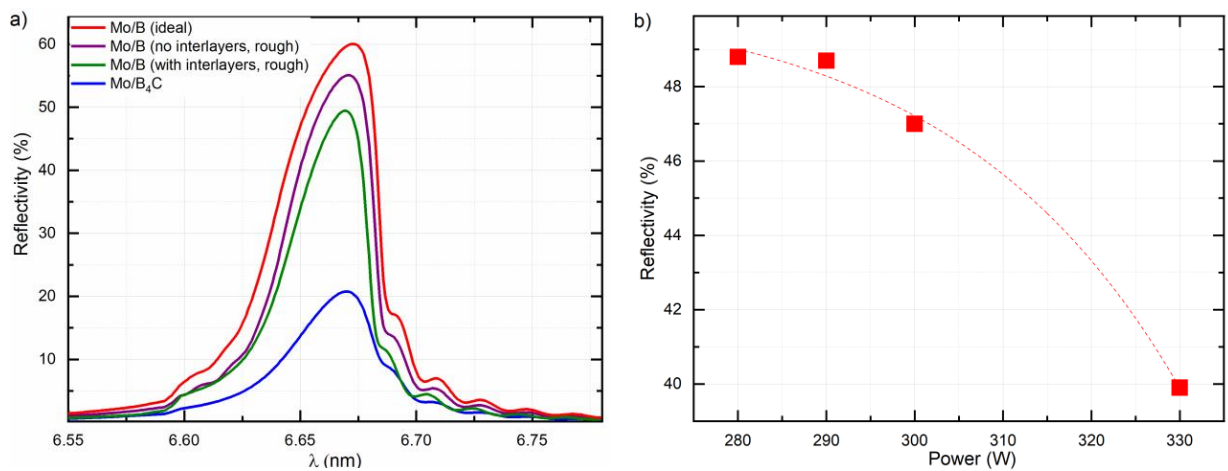

**Figure 6.** Calculated BEUV reflectivity at a working wavelength of 6.7 nm a) Comparison of the calculated peak reflectivity of different PMM models: ideal Mo/B; Mo/B with interface roughness; real Mo/B with interlayers; real Mo/B$_4$C. The angle of incidence was 5° off normal. b) Prediction of the reflectivity as a function of the B magnetron power based on the models derived from the GIXR simulations (Table 1, Supplementary Tables S1-S3).

It was demonstrated earlier that there is a good correlation between the real reflectivity of PMMs measured at synchrotron facilities and computer simulations performed based on GIXR models [14-16, 19, 34]. The reflectivity for the wavelength range of 6.5-6.8 nm was calculated for the PMM model described in Table 1. The simulation results indicated that the Mo/B PMM will exhibit a reflectivity of approximately 49 % that was about two times higher than that of the Mo/B$_4$C or Sb/B$_4$C PMMs [14, 34] and close to that of La(N)/B multilayers [16]. The B/La/LaN multilayers exhibit a high reflectivity of approximately 64 % [20]; however, their synthesis is expensive because it involves a highly sophisticated mechanism of deposition. Therefore, Mo/B PMMs are a cost-effective alternative to the B/La/LaN multilayers despite their relatively low reflectivity. A further investigation into the role of the B magnetron power could conveniently increase the performance of other PMMs like the La/B PMMs.

The B magnetron power significantly affected the structure of the PMMs; furthermore, it also affected the reflectivity of the PMMs in the BEUV range. The data from the simulation revealed that the temperature of the substrate increased to 120 °C with the increase in the B magnetron power to 330 W. This induced the decrease in the reflectivity to 40 %. The effect of the increase in the temperature during the deposition on the thickness of the interlayers was not included in the simulation. This factor would have further lowered the reflectivity to a significant extent.

The curvature measurements indicated that the residual stress of the PMMs was 0.8–0.9 GPa. Compressive stresses are commonly formed in amorphous coatings that are deposited at low temperatures; however, relatively low stress was observed in the coatings in this experiment. The compressive stresses in the Mo/B$_4$C and Pd/B$_4$C PMMs [14, 18] were several times higher than those in the Mo/B PMMs. The decrease in the mechanical stress in the Mo/B PMMs plays a crucial role in minimizing the risk of the delamination of the PMMs and deformation of the thin substrates

## 4. Conclusions

In this study, Mo/B PMMs were deposited by magnetron sputtering, and their structure was analyzed. The Mo/B PMMs comprised of smooth, amorphous Mo and B layers separated by ~0.4 nm-thick interlayers consisting of a mixture of MoB and MoB$_2$. Interlayers had an average density close to that of MoB$_2$. Then, the construction of optimal Mo/B PMM was designed, assuming the real nanostructure of Mo-B interfaces. The computer simulation of the optimal design indicated that the Mo/B PMMs will exhibit a reflectivity of approximately 50 % at a working wavelength of 6.7 nm.

It was shown that the structure and properties of the PMMs were sensitive to deposition temperature; therefore, the precise control of this parameter could significantly enhance the performance of the Mo/B PMMs. Furthermore, the results of this study can be extended to other boron-based PMMs.

**CRediT author statement**

**Oleksiy V. Penkov**: Conceptualization, Investigation, Resources. Writing - original draft, Writing - review & editing, Supervision, Project administration, Funding acquisition. **Igor A. Kopylets**: Conceptualization, Investigation, Resources, Validation, Writing - review & editing. **Valeriy V. Kondratenko**: Conceptualization, Resources, Validation, Writing - review & editing. **Mahdi Khadem**: Investigation, Resources, Writing - review & editing.

**Acknowledgment**

This work was supported by the Zhejiang University/University of Illinois at the Urbana-Champaign Institute and supervised by Oleksiy V. Penkov. M. Khadem was supported by the Brain Korea 21 Plus Project in 2020.

## References

[1] A. González, Trends in Processor Architecture in Harnessing Performance Variability / Embedded and High-performance Many/Multi-core Platforms, Springer International Publishing, Cham, 2019, 23-42   doi:10.1007/978-3-319-91962-1_2.

[2] Y. Sun, N.B. Agostini, S. Dong, D. Kaeli, Summarizing CPU and GPU Design Trends with Product Data, arXiv preprint arXiv:1911.11313  (2019) .

[3] V. Bakshi, H. Mizoguchi, T. Liang, A. Grenville, J. Benschop, Special Section Guest Editorial: EUV Lithography for the 3-nm Node and Beyond, J. Micro/Nanolith. MEMS MOEMS 16 (2017) 1, doi:10.1117/1.JMM.16.4.041001.

[4] N. Fu, Y. Liu, X. Ma, Z. Chen, EUV lithography: state-of-the-art review, J. Microelectron. Manuf. 2 (2019) 1, doi:10.33079/jomm.19020202.

[5] T. Otsuka, B. Li, C. O'Gorman, T. Cummins, D. Kilbane, T. Higashiguchi, N. Yugami, W. Jiang, A. Endo, P. Dunne, G. O'Sullivan, A 6.7-nm beyond EUV source as a future lithography source, SPIE Advanced Lithography, San Jose, California, (2012 ) 832214. doi:10.1117/12.916351.

[6] E. Louis, A.E. Yakshin, T. Tsarfati, F. Bijkerk, Nanometer interface and materials control for multilayer EUV-optical applications, Prog. Surf. Sci. 86 (2011) 255, doi:10.1016/j.progsurf.2011.08.001.

[7] Q. Huang, V. Medvedev, R. van de Kruijs, A. Yakshin, E. Louis, F. Bijkerk, Spectral tailoring of nanoscale EUV and soft x-ray multilayer optics, Appl. Phys. Rev. 4 (2017) 011104, doi:10.1063/1.4978290.

[8] B. Yu, C. Jin, S. Yao, C. Li, Y. Liu, F. Zhou, B. Guo, H. Wang, Y. Xie, L. Wang, Low-stress and high-reflectance Mo/Si multilayers for extreme ultraviolet lithography by magnetron sputtering deposition with bias assistance, Appl. Optics 56 (2017) 7462, doi:10.1364/AO.56.007462.

[9] S.S. Sakhonenkov, E.O. Filatova, A.U. Gaisin, S.A. Kasatikov, A.S. Konashuk, R.S. Pleshkov, N.I. Chkhalo, Angle resolved photoelectron spectroscopy as applied to X-ray mirrors: an in depth study of Mo/Si multilayer systems, Phys. Chem. Chem. Phys. 21 (2019) 25002, doi:10.1039/c9cp04582a.

[10] C. Montcalm, P.A. Kearney, J.M. Slaughter, B.T. Sullivan, M. Chaker, H. Pépin, C.M. Falco, Survey of Ti-, B-, and Y-based soft x-ray–extreme ultraviolet multilayer mirrors for the 2- to 12-nm wavelength region, Appl. Optics 35 (1996) 5134, doi:10.1364/AO.35.005134.

[11] O.V. Penkov, I.A. Kopylets, M. Khadem, T. Qin, X-Ray Calc: A software for the simulation of X-ray reflectivity, SoftwareX 12 (2020) 100528, doi:10.1016/j.softx.2020.100528.

[12] A.V. Vinogradov, Multilayer X-ray optics, Quantum Electron. 32 (2002) 1113, doi:10.1070/QE2002v032nl2ABEH002354.

[13] T.W. Barbee, Multilayer X-ray optics, Opt. Eng. 25 (1986) 899.

[14] M. Barthelmess, S. Bajt, Thermal and stress studies of normal incidence Mo/B4C multilayers for a 6.7 nm wavelength, Appl. Optics 50 (2011) 1610, doi:10.1364/AO.50.001610.

[15] P. Naujok, S. Yulin, A. Bianco, N. Mahne, N. Kaiser, A. Tünnermann, La/B4C multilayer mirrors with an additional wavelength suppression, Opt. Express 23 (2015) 4289, doi:10.1364/OE.23.004289.

[16] I.A. Makhotkin, E. Zoethout, R. van de Kruijs, S.N. Yakunin, E. Louis, A.M. Yakunin, V. Banine, S. Müllender, F. Bijkerk, Short period La/B and LaN/B multilayer mirrors for ~6.8 nm wavelength, Opt. Express 21 (2013) 29894, doi:10.1364/OE.21.029894.

[17] S.L. Nyabero, R.W. van de Kruijs, A.E. Yakshin, E. Zoethout, G. von Blanckenhagen, J. Bosgra, R.A. Loch, F. Bijkerk, Interlayer growth in Mo/B4C multilayered structures upon thermal annealing, J. Appl. Phys. 113 (2013) 144310, doi:10.1063/1.4800910.

[18] C. Morawe, R. Supruangnet, J.C. Peffen, Structural modifications in Pd/B4C multilayers for X-ray optical applications, Thin Solid Films 588 (2015) 1, doi:10.1016/j.tsf.2015.04.037.

[19] I. Kopylets, O. Devizenko, E. Zubarev, V. Kondratenko, I. Artyukov, A. Vinogradov, O. Penkov, Short-Period Multilayer X-ray Mirrors for "Water" and "Carbon Windows" Wavelengths, J. Nanosci. Nanotechnol. 19 (2019) 518, doi:10.1166/jnn.2019.16471.

[20] D.S. Kuznetsov, A.E. Yakshin, J.M. Sturm, R.W. van de Kruijs, E. Louis, F. Bijkerk, High-reflectance La/B-based multilayer mirror for 6.x nm wavelength, Opt. Lett. 40 (2015) 3778, doi:10.1364/OL.40.003778.

[21] B.L. Henke, E.M. Gullikson, J.C. Davis, X-Ray Interactions: Photoabsorption, Scattering, Transmission, and Reflection at E = 50-30,000 eV, Z = 1-92, Atom. Data Nucl. Data 54 (1993) 181, doi:10.1006/adnd.1993.1013.

[22] A. Herrera-Gomez, M. Bravo-Sanchez, O. Ceballos-Sanchez, M.O. Vazquez-Lepe, Practical methods for background subtraction in photoemission spectra, Surf. Interface Anal. 46 (2014) 897, doi:10.1002/sia.5453.

[23] Y. Zhang, Y.p. Zhao, Applicability range of Stoney's formula and modified formulas for a film/substrate bilayer, J. Appl. Phys. 99 (2006) 053513, doi:10.1063/1.2178400.

[24] A.D. Korotaev, A.N. Tyumentsev, Amorphization of metals by ion implantation and ion mixing methods, Russ. Phys. J. 37 (1994) 703, doi:10.1007/BF00559864.

[25] R. Benedictus, A. Böttger, E.J. Mittemeijer, Thermodynamic model for solid-state amorphization in binary systems at interfaces and grain boundaries, Phys. Rev. B 54 (1996) 9109, doi:10.1103/PhysRevB.54.9109.

[26] Y.G. Chen, B.X. Liu, Amorphous films formed by solid-state reaction in an immiscible Y–Mo system and their structural relaxation, Appl. Phys. Lett. 68 (1996) 3096, doi:10.1063/1.116434.

[27] P. Pershyn Yu., Y. Devizenko A., N. Zubarev E., V. Kondratenko V., L. Voronov D., M. Gullikson E., Scandium-silicon Multilayer X-ray Mirrors with CrB2 Barrier LayersScSiCrB, J. Nano Elec. Phys. 10 (2018) 05025, doi:10.21272/jnep.10(5).05025.

[28] J. Zhu, B. Ji, H. Jiang, J. Zhu, S. Zhu, M. Li, J. Zhang, Interface study of Sc/Si multilayers, Appl. Surf. Sci. 515 (2020) 146066, doi:10.1016/j.apsusc.2020.146066.

[29] D.L. Voronov, E.N. Zubarev, V.V. Kondratenko, Y.P. Pershin, V.A. Sevryukova, Y.A. Bugayev, Study of fast diffusion species in Sc/Si multilayers by W-based marker analysis, Thin Solid Films 513 (2006) 152, doi:10.1016/j.tsf.2006.01.070.

[30] R. Escamilla, E. Carvajal, M. Cruz-Irisson, F. Morales, L. Huerta, E. Verdin, XPS study of the electronic density of states in the superconducting Mo2B and Mo2BC compounds, J. Mater. Sci. 51 (2016) 6411, doi:10.1007/s10853-016-9938-z.

[31] R. Jothi Palani, Y. Zhang, P. Scheifers Jan, H. Park, P.T. Fokwa Boniface, Molybdenum diboride nanoparticles as a highly efficient electrocatalyst for the hydrogen evolution reaction, Sustainable Energy Fuels 1 (2017) 1928, doi:10.1039/C7SE00397H.

[32] H. Park, A. Encinas, J.P. Scheifers, Y. Zhang, B.P. Fokwa, Boron-Dependency of Molybdenum Boride Electrocatalysts for the Hydrogen Evolution Reaction, Angew. Chem. Int. Ed. 56 (2017) 5575, doi:10.1002/anie.201611756.

[33] D. Kruger, A. Penkov, Y. Yamamoto, A. Goryachko, B. Tillack, D. Krüger, Characterization of Ge gradients in SiGe HBTs by AES depth profile simulation, Appl. Surf. Sci. 224 (2004) 51, doi:10.1016/j.apsusc.2003.08.027.

[34] I.A. Kopylets, V.V. Kondratenko, E.N. Zubarev, D.L. Voronov, E.M. Gullikson, E.A. Vishnyakov, E.N. Ragozin, Fabrication and characterization of Sb/B4C multilayer mirrors for soft X-rays, Appl. Surf. Sci. 307 (2014) 360, doi:10.1016/j.apsusc.2014.04.038.